# ETROC1: The First Full Chain Precision Timing Prototype ASIC for CMS MTD Endcap Timing Layer Upgrade

**X. Huang,[a,1] Q. Sun,[b] D. Gong,[b] P. Gwak,[c] D. Kim,[d] J. Lee,[e] C. Liu,[f] T. Liu,[f] T. Liu,[b] S. Los,[b] S. Miryala,[b] S. Nanda,[e] J. Olsen,[b] H. Sun,[f] J. Wu,[b] J. Ye,[f] Z. Ye,[e,g] L. Zhang,[f] and W. Zhang[f]**

[a] *University of California, Santa Barbara,*
*Santa Barbara, CA 93106, USA*

[b] *Fermi National Accelerator Laboratory,*
*Batavia, IL 60510, USA*

[c] *Chonnam National University,*
Gwangju, 61186, Korea

[d] *Kansas State University,*
*Manhattan, KS 66506, USA*

[e] *University of Illinois Chicago,*
*Chicago, IL 60607, USA*

[f] *Southern Methodist University,*
*Dallas, TX 75275, USA*

[g] *Lawrence Berkeley National Laboratory,*
*Berkeley, CA 94720, USA*
*E-mail*: xinghuang@ucsb.edu

ABSTRACT: we present the design and characterization of the first full chain precision timing prototype ASIC, named ETL Readout Chip version 1 (ETROC1) for the CMS MTD endcap timing layer (ETL) upgrade. The ETL utilizes Low Gain Avalanche Diode (LGAD) sensors to detect charged particles, with the goal to achieve a time resolution of 40 – 50 ps per hit, and 30 – 40 ps per track with hits from two detector layers. The ETROC1 is composed of a 5 × 5 pixel array and peripheral circuits. The pixel array includes a 4 × 4 active pixel array with an H-tree shaped network delivering clock and charge injection signals. Each active pixel is composed of various components, including a bump pad, a charge injection circuit, a pre-amplifier, a discriminator, a digital-to-analog converter, and a time-to-digital converter. These components play essential roles as the front-end link in processing LGAD signals and measuring timing-related information. The peripheral circuits provide clock signals and readout functionalities. The size of the ETROC1 chip is 7 mm × 9 mm. ETROC1 has been fabricated in a 65 nm CMOS process, and extensively tested under stimuli of charge injection, infrared laser, and proton beam. The time resolution of bump-bonded ETROC1 + LGAD chipsets reaches 42 – 46 ps per hit in the beam test.

---

[1] Corresponding author.

## Contents

## 1. Introduction

The Large Hadron Collider (LHC) at CERN, Switzerland has been under upgrade to deliver higher collision rates. The instantaneous luminosity of the upgraded LHC, namely High-Luminosity LHC (HL-LHC) [1], will increase by a factor ~5, reaching $5.0 \times 10^{34}$ $cm^{-2}s^{-1}$ in the initial phase, and up to $7.5 \times 10^{34}$ $cm^{-2}s^{-1}$ later. In the initial phase, there will be an average of 140 proton-proton collisions (pile-up) in a bunch crossing. In the later phase, this number will increase to around 200 pile-up collisions per bunch crossing. To cope with the increased challenges posed by the higher luminosity and pile-up rate of the HL-LHC, the CMS collaboration has decided to upgrade its detector. One specific upgrade is the addition of a new MIP (Minimum Ionizing Particle) Timing Detector (MTD) [2]. This new detector will help maintain the CMS detector's excellent performance by providing accurate timing information for particles passing through it, which is crucial for distinguishing particles from different pile-up collisions to minimize the impact of the enhanced pile-up rate. Overall, the HL-LHC and the subsequent CMS detector upgrade represent the ongoing efforts to push the boundaries of particle physics research by increasing the collision rate and enhancing the capabilities of the experimental detectors. These upgrades enable scientists to explore new physics phenomena and gather more precise data to further our understanding of the fundamental building blocks of the universe.

The CMS MTD will be separated into the Barrel Timing Layer (BTL), covering the pseudo rapidity region $|\eta| < 1.45$, and the Endcap Timing Layer (ETL), covering the forward region $1.6 < |\eta| < 3.0$. The BTL will be instrumented with Lutetium–Yttrium Orthosilicate (LYSO) [3] crystal scintillators coupled with Silicon Photomultipliers (SiPMs), covering a total active area of 38 $m^2$. The SiPMs will be read out by the Time-of-Flight at High-Rate (TOFHIR) ASIC [4] [5]. The ETL will be composed of 2 disks in each endcap region (4 disks in total) with Low-Gain Avalanche Diode (LGAD) sensors [6] to cover a total active area of 14 $m^2$. The LGAD sensors will be read out by the ETL Readout Chip (ETROC) ASIC. The ETL is designed to provide precision timing measurements for charged tracks with 30 – 40 ps resolution per track. The LGAD gain is 10 – 30 to maximum the signal to noise ratio. LGAD sensors from the HPK2 [7] and FBK UFSD3.2 [8] productions have been measured both in the laboratory and during beam tests, finding that they meet all the specifications. The expected Most Probable Value (MPV) of the charge from the LGAD sensor is approximately 15 fC, with an expected operating range of charge of 8 fC – 25 fC [9]. The radiation tolerance specification for Total Ionizing Dose (TID) is 100 Mrad [10].

The development of the ETROC ASIC is divided into three prototyping phases. ETROC0 [11] consists of a single channel analog front-end with a charge injection circuit, a pre-amplifier, and a discriminator. ETROC0 chips wire bonded to LGAD sensors have been demonstrated to provide a time resolution of around 33 ps from the pre-amplifier waveform analysis and around 42 ps from the discriminator pulse analysis in beam test. ETROC1 [12] is the first full chain precision timing prototype. It has a 4 × 4 array of active pixels, each of which includes the above mentioned analog front-end and a new Time-to-Digital Converter (TDC) [13] for time-of-arrival (TOA) and time-over-threshold (TOT) measurements. ETROC1 aims to study and demonstrate the performance of the full LGAD signal processing chain, with the goal of achieving 40 – 50 ps time resolution per hit (30–40 ps per track with hits from two detector layers). ETROC1 has been fabricated in a 65 nm Complementary Metal-Oxide-Semiconductor (CMOS) process, and extensively tested under stimuli of charge injection, laser, and beam. ETROC2 is the first full size (16 × 16 pixels) and full functionality prototype. Several ETROC2 building blocks [13] - [18] were taped out and fully verified before the ETROC2 submission in October 2022. ETROC2 has been fabricated and is under test. Concurrently, the similar purpose and scope timing ASIC ALTIROC (ATLAS LGAD Time Read Out Chip) for ATLAS upgrade is under development [19].

In this paper, we describe the ETROC1 design including the new TDC and clock distribution network, and then focus on the bench test and beam test results of bump-bonded ETROC1 + LGAD chipsets, and how the knowledge gained from bump-bonded ETROC1 testing is applied to guide the ETROC2 design. Section 2 describes the ETROC1 design and structure. Section 3 presents the full test results with charge injection, laser, and beam. Section 3 also analyses noise related to a 40 MHz clock and proposes solutions to address this issue. Section 4 draws the conclusion.

## 2. Design of ETROC1

### 2.1 Overall structure

ETROC1 is the first full chain precision timing prototype of the ETROC series. The diagram of ETROC1 is shown in **Figure 1** (left), including a 5 × 5-pixel matrix with pixel size of 1.3 × 1.3 $mm^2$, and periphery circuits. The 5 × 5-pixel matrix includes a 4 × 4 active pixel matrix (in orange) with a precision clock distribution network (in green) that is scalable to the final full size 16 × 16 array, a single pixel for test purposes (in red), and 8 dummy pixels (in blue). The 8 dummy pixels provide electrical and mechanical support to the sensor. Signal pads of dummy pixels are pulled down. 17 (16 for the active pixel array and 1 for the standalone pixel) out of the 25 pixels are connected to the readout circuits. A 3 paralleled H-tree shaped network (green trace in **Figure 1** (left)) based on inverter is employed to deliver the TDC 40 MHz clock, TDC reference strobe signal, and charge injection signal. The periphery circuits consist of an inter-integrated circuit (I2C), a clock generator, a simple readout (SRO) module, a diagnostic mode readout (DMRO) module, and several line drivers (eTxes) [20] and line receivers (eRxes) [20]. The standalone TDC is a test block placed to study the performance of the TDC. A generic I2C slave is used twice in ETROC1. Each slave provides 32 bytes for reading and 16 bytes for writing by ETROC1. A 4-bit chip ID and a 4-bit chip reversion are available as well. The registers in the I2C slave are triplicated to mitigate single event upset.

The diagram of the active pixel is shown in **Figure 1** (right). Each active pixel is composed of a bump pad, a charge injection circuit, a pre-amplifier (PA), a discriminator, a 10-bit digital-to-analog converter (DAC), and a TDC. The charge injection circuit is added to emulate charge signals from LGAD sensors to characterize the front-end circuits without sensor and irradiation. The pre-amplifier converts the charge signal to a voltage signal with a fast leading-edge to make it suitable for further processing and digitization. The study of the capacitor at the input of the pre-amplifier to externally mimic the sensor has been done in ETROC0 [9]. No capacitor is added to mimic the sensor in ETROC1. The discriminator digitizes the analog voltage signal produced by the pre-amplifier by comparing the amplitude of this signal to a threshold voltage provided by the 10-bit Digital-to-Analog Converter (DAC). The TDC is responsible for measuring timing information related to the detected signal.

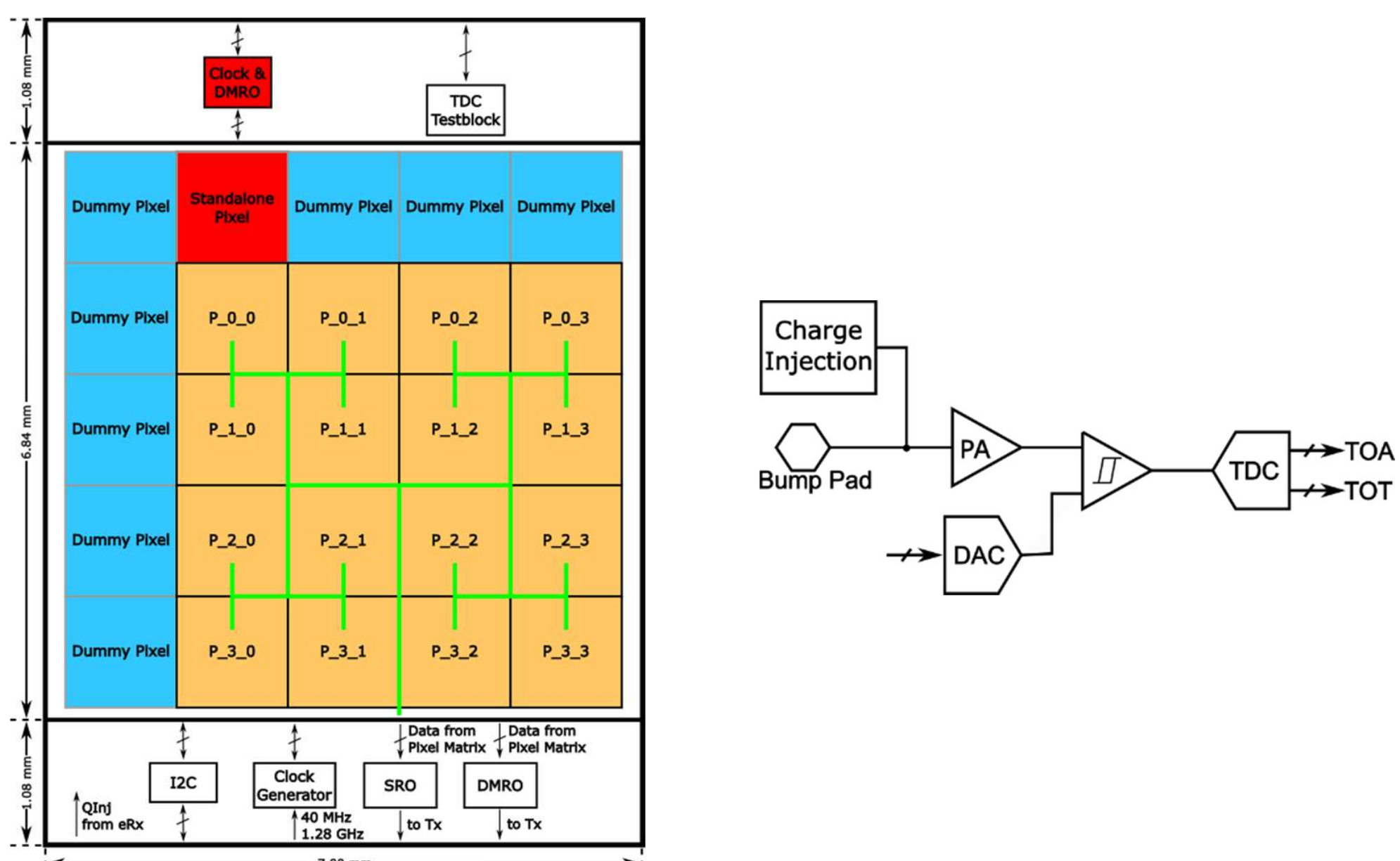

**Figure 1.** Diagram of the overall ETROC1 chip (left) and an active pixel (right).

### 2.2 Charge injection

The charge injection circuit injects charge to the pre-amplifier for testability. The diagram of the charge injection circuit is shown in **Figure 2**. It takes a step pulse with a known voltage amplitude (QInj) to generate charge on the capacitor (C) connected to the pre-amplifier. A 5-bit control bit connected to I2C selects the injected charge from 1 fC to 32 fC. The charge injector is enabled by turning on the switch (SW1) between the capacitor and the pre-amplifier and is triggered by the charge injection pulses from an off-chip pulse generator that are distributed to the pixel array by the clock H-tree.

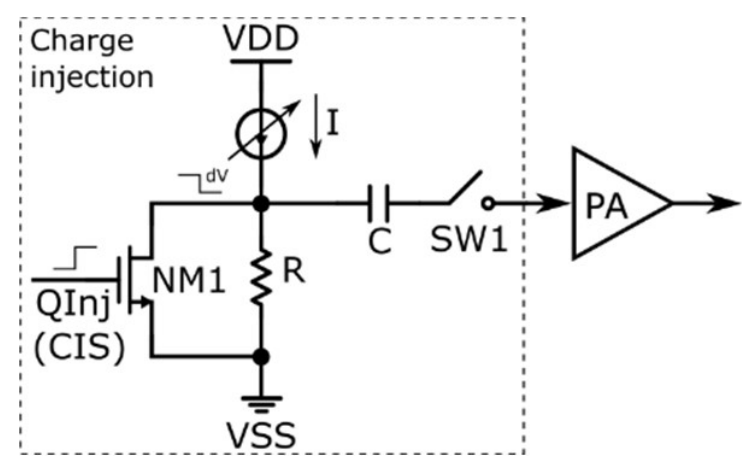

**Figure 2.** Diagram of the charge injection circuit in ETROC1 pixel.

### 2.3 Pre-amplifier

The schematic of the pre-amplifier is shown in **Figure 3**. The pre-amplifier is a buffered transimpedance amplifier (TIA) [11]. The main amplified stage of the TIA is formed by transistors NM3, NM2, PM4 and PM3 with the feedback resistor network (NM6, NM7, R4, R5, and R6). The feedback resistor network between 4.4 kΩ and 20 kΩ is used to adjust the gain and fall time of the pre-amplifier to compensate the possible effect of radiation damage. A current source (I) and transistors (PM1, PM2, and NM1) provide the bias voltages for PM3 and NM2, respectively. A bias branch consisting of PM5, R1 and R2, provides adjustable current for the input transistor, NM3, to boost its transconductance. A capacitance network (NM4 and C) is programmable to tune the rise time and bandwidth. The output stage is a PMOS source follower (PM6 and R3) to isolate the noise from the discriminator and provides reasonable baseline for the output signal.

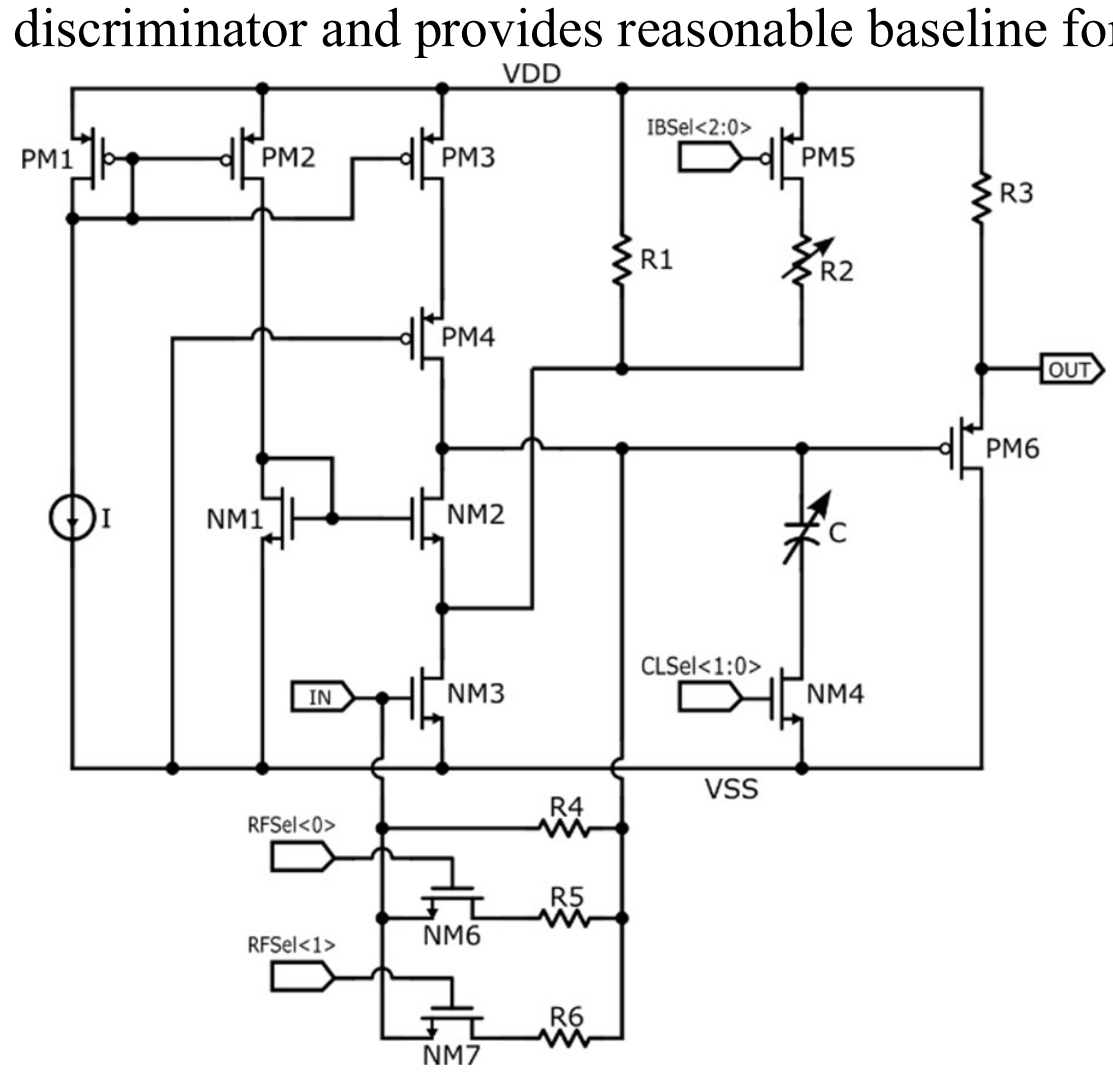

**Figure 3.** Schematic of the pre-amplifier in ETROC1 pixel.

## 2.4 Discriminator

The discriminator is used to convert analog pulses from the pre-amplifier to digital signals. The schematic of the discriminator is shown in **Figure 4**. It consists of a three-stage pre-amplifier and a comparator with programmable hysteresis. The three-stage pre-amplifier (R1 ~ R6, NM1 ~ NM6, and I1 ~ I3) in the discriminator is employed to amplify the small input pulses to a level where they can be reliably processed by the subsequent comparator. The comparator (PM1 ~ PM3, PM6 ~ PM8, NM7 ~ NM10, and I4) digitizes the differential input at the crossing point with an adjustable hysteresis (PM4, PM5, PM9, and PM10) ranging from 0 to 1 mV. The hysteresis helps improve noise immunity and prevent false triggering near the threshold. A buffer following the OUT node helps maintain a low-impedance output from the comparator while driving the following stages without distortion.

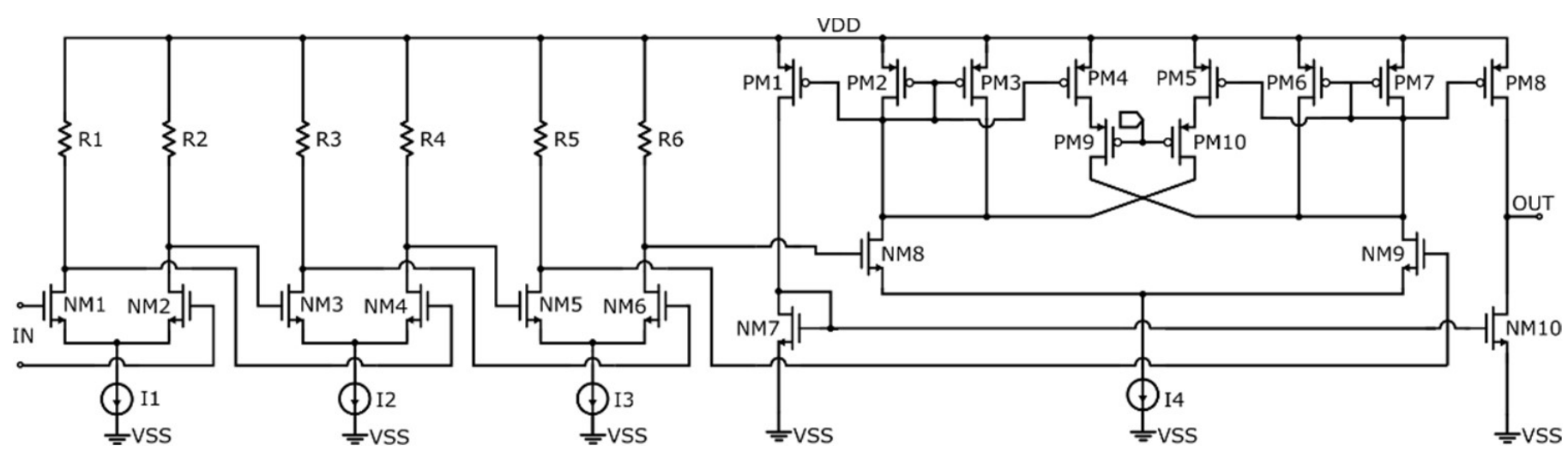

**Figure 4.** Schematic of the discriminator in ETROC1 pixel.

A 10-bit DAC is included in the pixel to generate the threshold voltage of the discriminator. As the baseline of the pre-amplifier varies with temperature and bias setting, the DAC is designed to cover a range from 0.6 V to 1 V with a step size of 0.4 mV. The DAC employs a resistor-string structure to achieve monotone over its operation range. A noise filter is used to limit the noise of the threshold voltage below 0.1 mV to minimize the DAC noise contribution to the timing performance.

## 2.5 TDC

The TDC [13] is based on a simple delay-line approach without any control mechanisms, which is originally prototyped and implemented [21] in Field-Programmable Gate Array (FPGA) due to their flexibility and reconfigurability. The gate-level schematic of the TDC core is shown in **Figure 5**. One of the challenges of the ETROC design is that the TDC is required to consume less than 200 μW for each pixel at the nominal hit occupancy of 1%. To meet the low-power requirement, a single delay line without the need of delay locked loop (DLL) to control individual delay cell is used with a cyclic structure to reduce the number of delay cells. The TDC measures TOA and TOT at the same time. A double-strobe self-calibration scheme implemented to compensate for variations in process, power supply voltage, and temperature (PVT). Each hit (or event) is registered twice at two consecutive clock rising edges with an accurate time difference of the known 320 MHz clock period. The duplicated registration serves the purpose of refining timing measurements and enhancing accuracy to mitigate the effect of PVT variations. The data width of the TDC output is 30 bits including 10-bit TOA code, 9-bit TOT code, 10-bit calibration code, and 1-bit hit flag. For ETL application, the occupancy is expected to be lower (1% level), and the extra calibration data does not affect the readout much, while providing more flexibility in calibration depending on the environmental changes such as power supply voltage, temperature, radiation damage, etc.

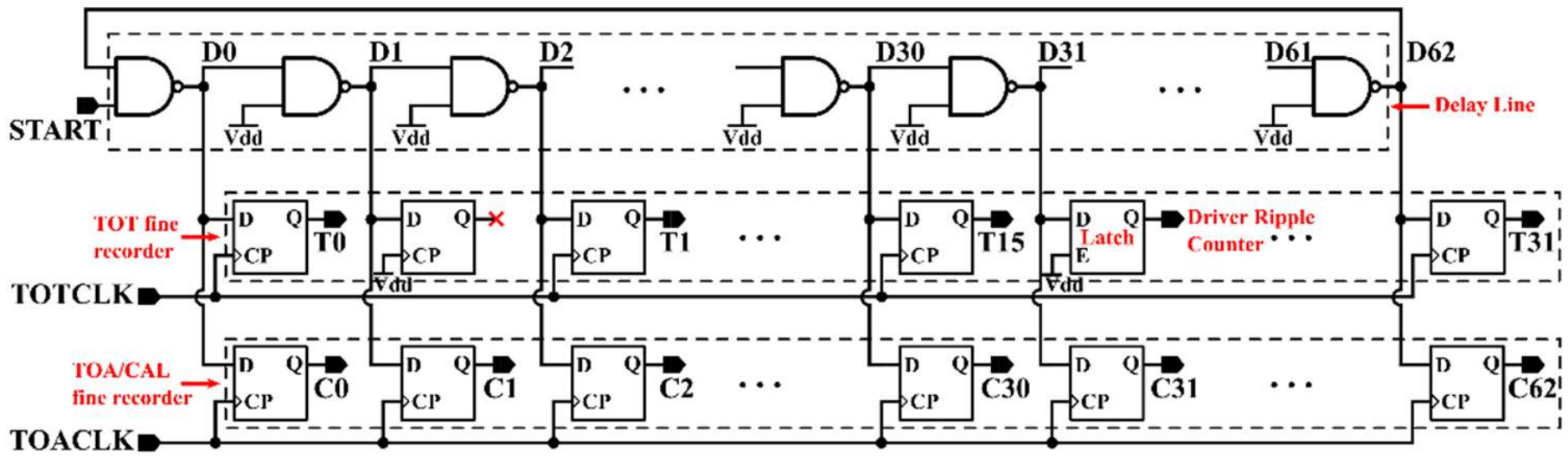

**Figure 5.** Gate-level schematic of the TDC core in ETROC1 pixel.

The overall performance of the TDC has been evaluated separately [13] and meets the CMS ETL upgrade requirements. To fully cover the entire range of the TDC, the input signal to be tested is a repeating pulse with an adjustable width at 40.001 MHz which is slightly different from the TDC clock (40 MHz). We utilize the time discrepancy of 0.625 ps to make the input pulse chase the TDC clock. The TOA has a bin size of 17.8 ps within its effective dynamic range of 11.6 ns. The TOT has a bin size of 35.4 ps within its measured dynamic range of 9.8 ns. The effective measurement precisions of the TDC are 5.6 ps and 9.9 ps for the TOA and 10.4 ps and 16.7 ps for the TOT with and without nonlinearity correction, respectively.

### 2.6 Clock system

The diagram of the clock system in ETROC1 is shown in **Figure 6**. ETROC1 receives a differential 1.28 GHz off-chip clock. This is a temporary option for testing purposes in ETROC1. For the final ASIC, the on-chip phase lock loop (PLL) [15] will be implemented that needs a differential 40 MHz clock from lpGBT to generate a 1.28 GHz clock. eRx converts the differential 1.28 GHz input clock to a single ended clock that is connected to a divider to produce a 40 MHz clock. A phase shifter based on a shared DLL is implemented to produce a 320 MHz clock and a programmable phase rotation (up to 360 degree) with a time resolution of 97.6 ps for both 40 MHz and 320 MHz clocks. A multiplexer (MUX1) is used to select the internal clocks or external clocks of 40 MHz and 320 MHz and then output to the TDC reference strobe generator and MUX2. The TDC reference strobe generator exploits an 8 D flip flop (DFF) chain to produce a mask signal to generate a TDC reference strobe signal which has two successive pulses at 320 MHz frequency. The TDC reference strobe signal is used to calibrate the timing measurement. The MUX2 is to select the clock combinations output to the clock H-tree or to output pads through eTx.

The TDC 40 MHz clock, TDC reference strobe signal, and charge injection pulse are delivered through a paralleled 4 × 4 H-shaped clock tree. An inverter-based buffering approach is used to balance the rising edge and the falling edge of the signal. The layout of inverters with enclosed layout transistor (ELT) is adopted from a lpGBT project [22] to improve the robustness against Total Ionizing Dose (TID). To eliminate the variations, modular wiring methodology is used in the layout with the basic cell of an inverter and a fixed length metal trace. The propagation delay is 0.3 ps. The skew is 1.0 ps. The jitter is 0.4 ps. The power consumption is 0.4 mW. The 4 × 4 clock H-tree can be scaled up to a 16 × 16 clock tree in the following iteration of ETROC design.

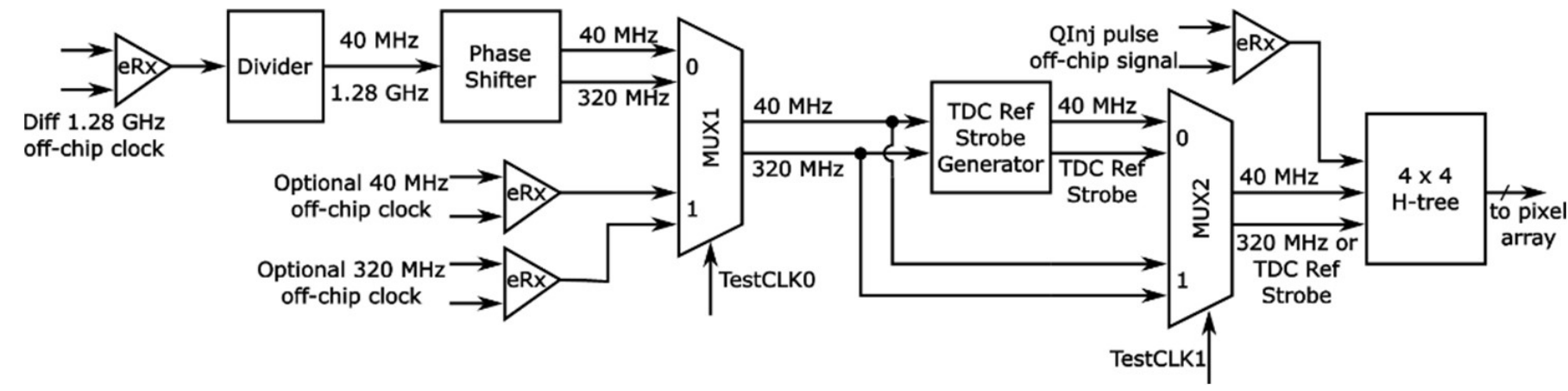

**Figure 6.** Diagram of the clock system in ETROC1.

## 2.7 Readout logic

ETROC1 has two readout modes, diagnostic mode readout (DMRO), and simple readout (SRO). The diagram of the DMRO is shown in **Figure 7** (left). The pixel array is clustered into four columns according to the physical location in the pixel array. The TDC data of each pixel is buffered with enable signal to the corresponding column data lane and then routed to a multiplexer (MUX1). A scrambler is utilized to scramble the 30-bit TDC data. Both the input and the output of the scrambler are connected to the input of MUX2. A 2-bit header is added to the output of the scrambler to form 32-bit data for the serializer at 40 MHz clock. A pseudo random binary sequence 7 (PRBS7) block is adopted for test purposes. The serial data at 1.28 Gbps is output to the pads via eTx to be read out.

The diagram of the SRO is shown in **Figure 7** (right). A circular buffer in each pixel stores data from TDC. Once an L1 acceptance is present, data in the circular buffers is frozen. All the data in the circular buffer of selected pixels is then read out at 1.28 Gbps with SRO block. Circular buffers accept TDC data again when all the data is delivered. The main difference between the DMRO and the SRO is that the DMRO can continuously deliver TDC data for a selected pixel while TDC data for one or more selected pixels could be stored in on-chip memory for readout later in the SRO.

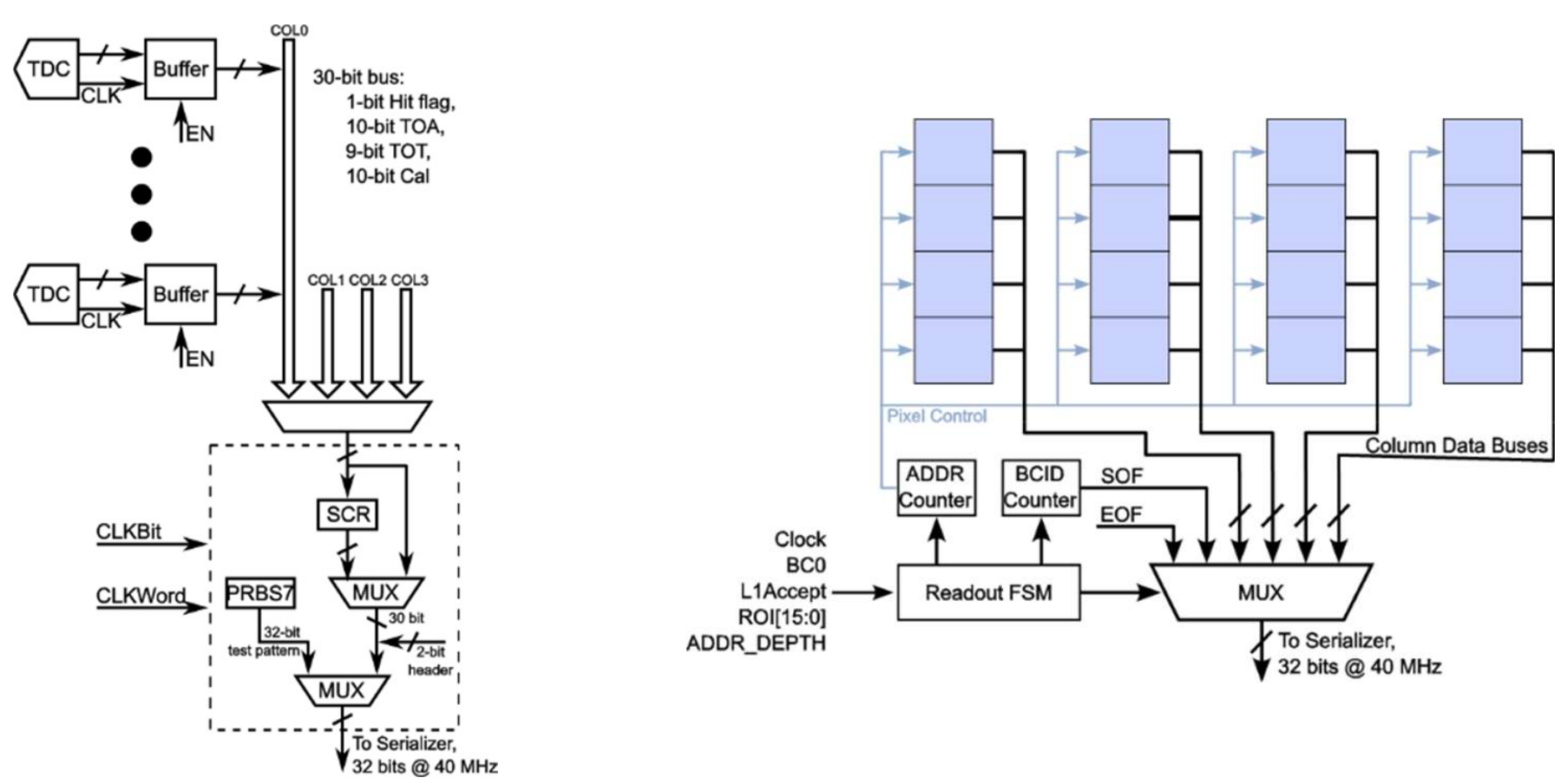

**Figure 7.** Diagram of the DMRO logic (left) and SRO (right) in ETROC1.

## 3. Test results

Bare ETROC1 chips (without LGAD sensor) and bump-bonded chipsets comprising ETROC1 chips and LGAD sensors have been tested extensively under different types of stimuli such as charge injection, infrared laser, and proton beam, respectively. These tests are essential for validating the functionality and performance of ETROC1.

A photograph of a bare ETROC1 chip is shown in **Figure 8** (left). The block diagram of the ETROC1 test setup is shown in **Figure 8** (middle). A photograph of an ETROC1 test setup in laboratory is shown in **Figure 8** (right). Three DC power supplies (Keysight Technologies E36311A) provide 1.2 V for ETROC1 test board. A clock generator (Hewlett Packard Model 8133A) provides 1.28 GHz clock to ETROC1. An FPGA (AMD Kintex7 FPGA KC705 Evaluation Kit) is used to buffer the 1.28-Gbps output data from ETROC1 and transmit the data to a personal computer (PC) through Ethernet cable for off-line analysis. Simultaneously, the FPGA provides I2C communication between ETROC1 and the PC.

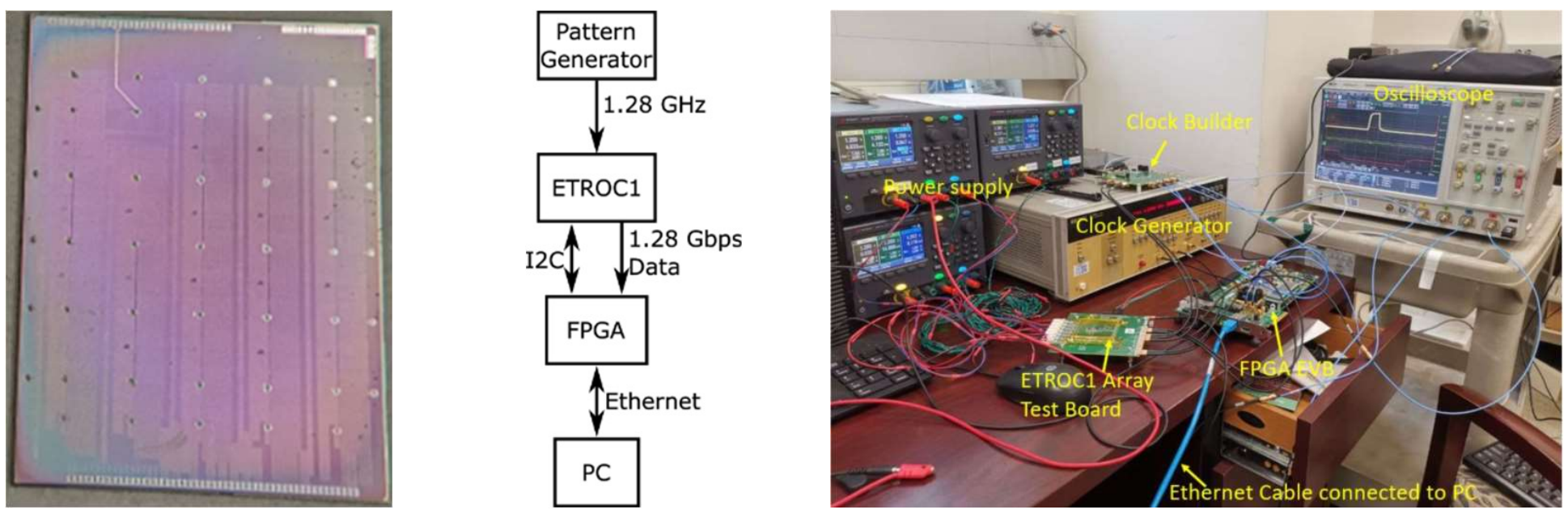

**Figure 8.** Photograph of ETROC1 chip (left), block diagram of ETROC1 test (middle), and picture of bare ETROC1 test setup (right).

### 3.1 Results of charge injection

The timing performance of the full signal processing chain as well as the 4x4 clock distribution network implemented in ETROC1 has been studied using the on-pixel charge injection feature. By injecting known charges and sweeping the threshold of the discriminator with DAC setting simultaneously, data is acquired from the dedicated pixel's output. The jitter of the TOA code as a function of the threshold (DAC) from a pixel of a bare ETROC1 chip is shown as an example in **Figure 9** (left). The corresponding charge threshold can be read from the rising edge of each line in this plot. The X-axis value of 500 stands for about 4 fC, the X-axis value of 550 represents about 8 fC, and so on. Here jitter refers to the standard deviation of the TOA code distribution (Y axis) concerning the TDC performance. The jitter performance of the 4 × 4-pixel array with the input charge of 15 fC and the threshold of 4 fC, after converting the TOA code to the TOA, is shown in **Figure 9** (right). As can be seen, the jitter ranges from 8.8 ps to 11.9 ps for this bare ETROC1 chip. The jitter for bump bonded ETROC1 + LGAD chipsets is about 20 ps. The resolution difference between the bare chip and ETROC1 + LGAD sensor is largely caused by the capacitor at the input of the preamplifier. In addition, the bump bonded ETROC1 has the 40 MHz noise issue, and that affects the resolution. For the 40 MHz noise issue, please refer to later subsection 3.4.

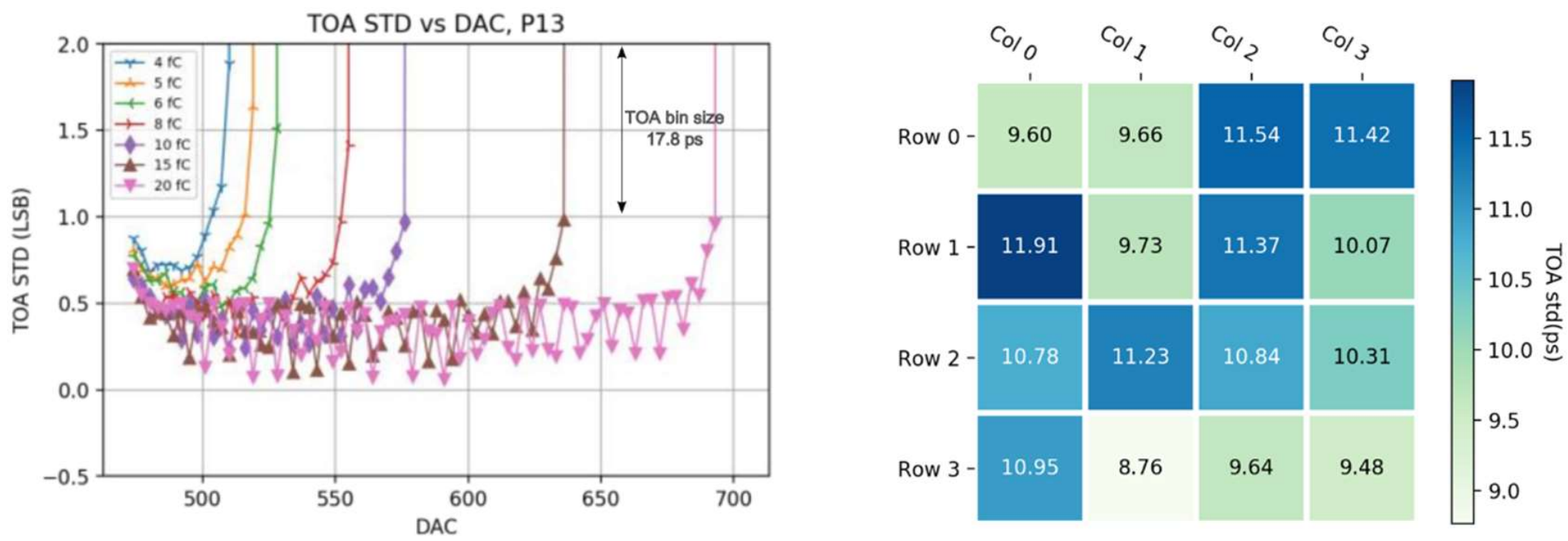

**Figure 9.** TOA results with charge injection feature for bare ETROC1.

### 3.2 Results of laser test

Bump-bonded ETROC1 + LGAD chipsets have been tested under infrared laser at Fermilab. The TOA and TOT were read out from ETROC1 with different laser intensity. The distribution of TOA versus TOT from such a measurement is shown in **Figure 10** (left). The analyzed jitter from the accumulated data before (in blue) and after (in orange) time walk correction is shown in **Figure 10** (right). The charge used in the laser test is 15 fC, the threshold of 8 fC is chosen due to the effect of the 40 MHz noise issue. The chipset has achieved a resolution of around 20 ps for TOA after time walk correction at large laser intensity (low laser attenuation).

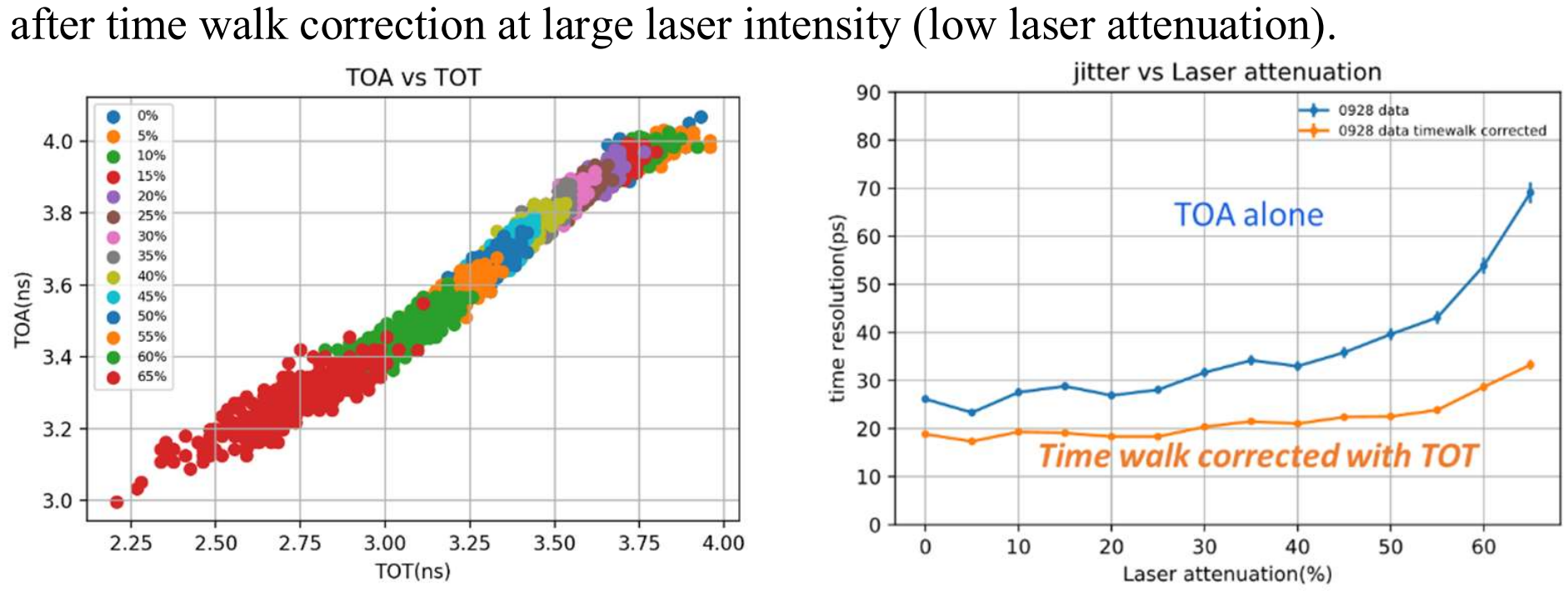

**Figure 10.** Test results with infrared laser stimulus.

### 3.3 Results of beam telescope

Bump-bonded ETROC1 + LGAD chipsets have also been tested with 120 GeV proton beam at the Fermilab Test Beam Facility (FTBF). The data acquisition system was the same as used in charge injection and laser tests. Different from a single test board under stimuli of charge injection and laser, three ETROC1 test boards were stacked in a row and formed a beam telescope, as shown in **Figure 11** (left). Upon receiving trigger signals from board B1, TDC data from all three boards containing timing information of the same proton tracks were recorded. Pictures of the beam test setup and telescope are shown in **Figure 11** (right). The charge used in the beam test is 15 fC, and the threshold of 8 fC is used because of the 40 MHz noise issue.

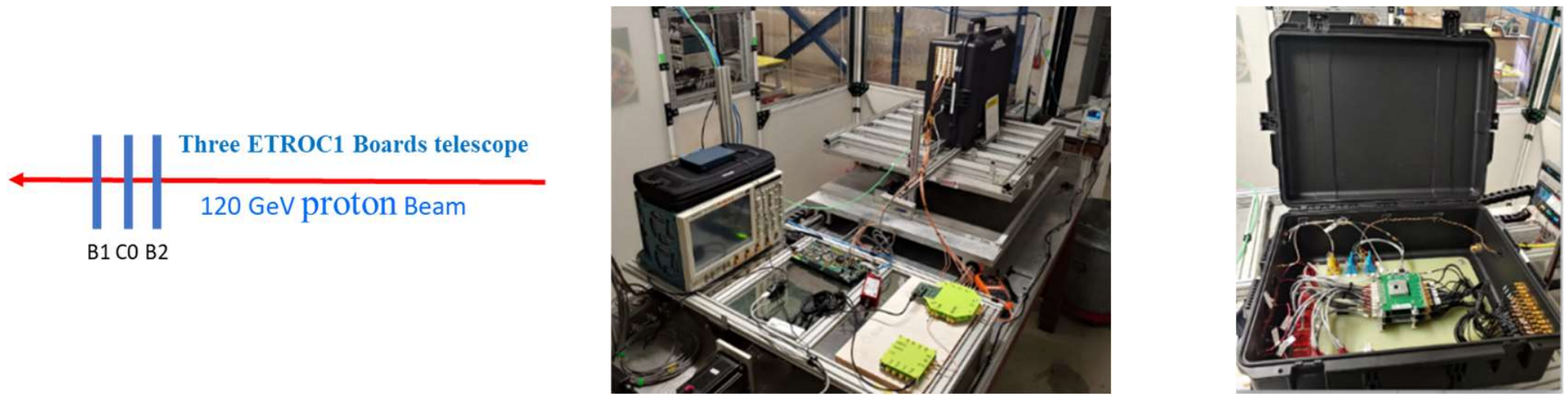

**Figure 11.** Diagram (left), test setup picture (middle), and telescope setup in the suitcase (right) for ETROC1 beam telescope.

Distributions of the difference in the recorded TOA between every two boards, after time walk correction based on the recorded TOT, are shown in **Figure 12**. The time resolution of each chipset is calculated as $\sigma_i = \sqrt{0.5 \cdot \left(\sigma_{ij}^2 + \sigma_{ik}^2 - \sigma_{jk}^2\right)}$, and is found to be between 42 – 46 ps from the 3 boards. By using the average value of the calibration code for each board during the data taking period, the time resolution can be improved to 41 – 43 ps, as shown in **Figure 13**. This is because the calibration code determination is subject to fluctuation. By using the average and thus fixed value of the calibration code, the contribution from the calibration code fluctuation to the time resolution can be removed.

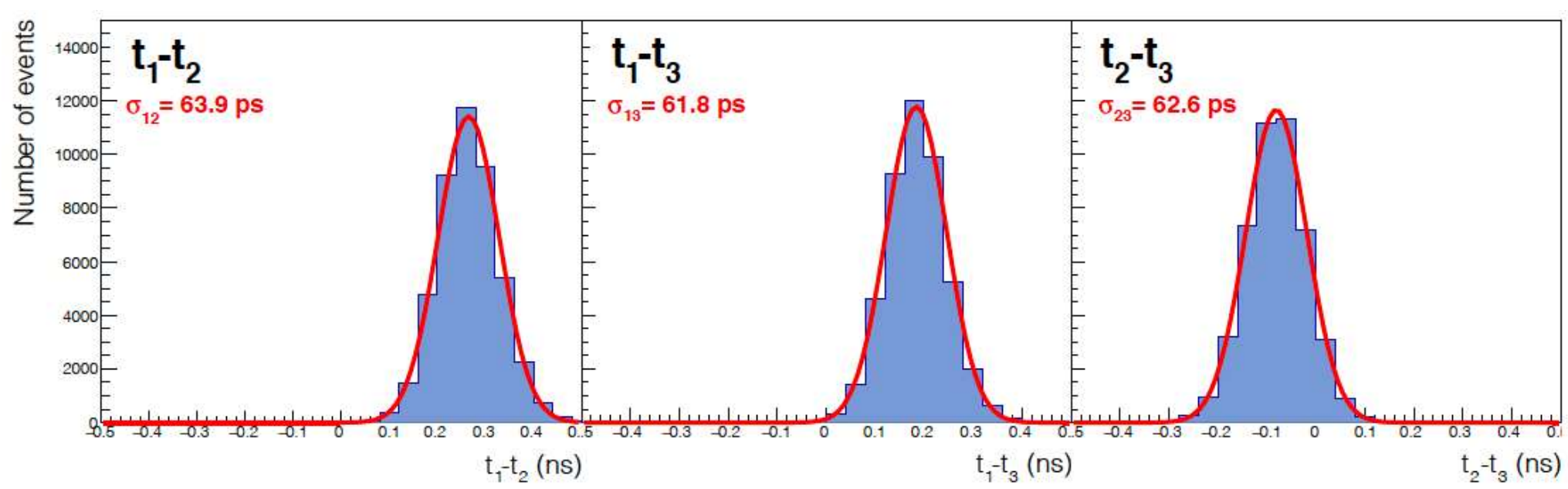

**Figure 12.** Test results with proton beam obtained with event-by-event calibration codes.

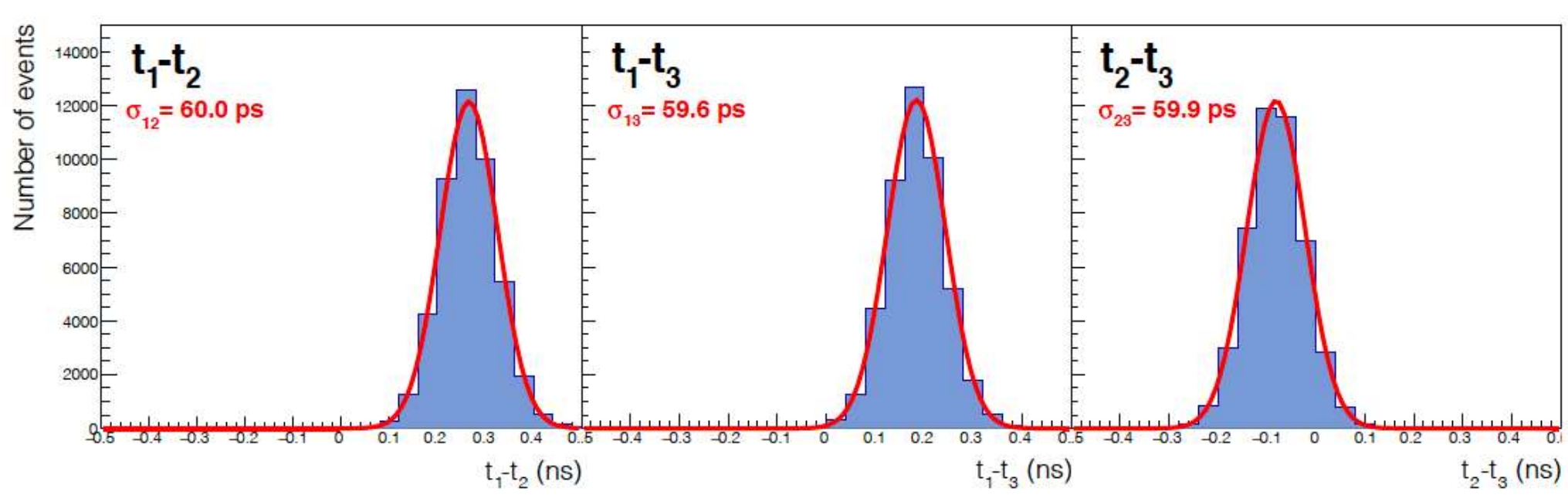

**Figure 13.** Test results with proton beam obtained with average values of the calibration codes.

### 3.4 Main lesson learned from ETROC1

After ETROC1 chips were bump-bonded with LGAD sensors, noise related to 40 MHz clock was observed in the discriminator output at low discriminator thresholds, as shown in **Figure 14** (left). The noise disappeared from the discriminator output when the threshold was increased high enough (e.g., DAC=420, corresponding to ~8 fC). The 40 MHz noise was also observed in the beam test as shown in **Figure 14** (right). In the TOT distribution, a pattern of the 40 MHz noise exists but lower than the proton beam signal. A pattern with phase difference is shown in the TOA distribution. Good time resolution was obtained from the 4x4 active pixels with high enough threshold (> 8 fC), as described in the previous section.

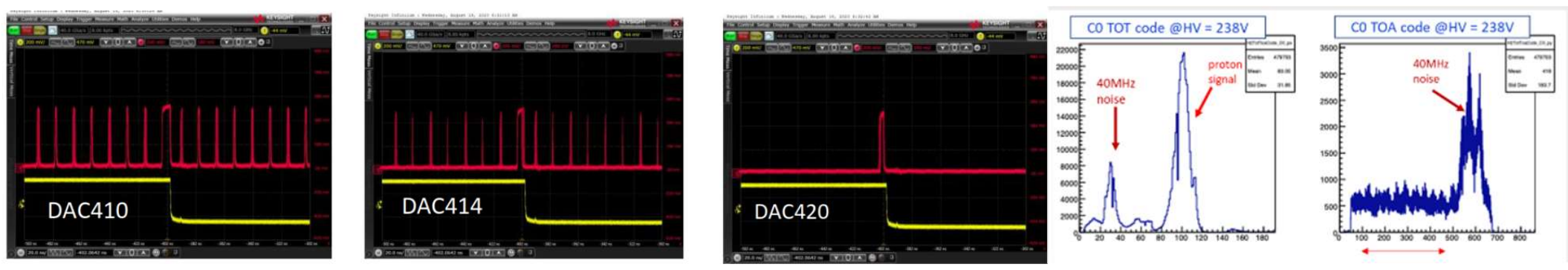

**Figure 14.** 40 MHz noise observed during charge injection (left) and beam test (right) from bump-bonded ETROC1 + LGAD chipsets.

Investigations have been conducted to identify the noise source. It has been found that bare ETROC1 chips do not have the 40 MHz noise issue, which is the benchmark of the noise analysis. The performance of the standalone pixel with the circular buffer disabled in ETROC1 chips (**Figure 1**) that were bump-bonded to LGAD sensors is very close to bare ETROC1 chips. A diagram of the bump bonded ETROC1 + LGAD chipset is shown in **Figure 15**. The following actions have been taken, 1) increasing the spacing with bump on both sides from 50 µm to 120 µm, 2) connecting only one guard ring pad of LGAD sensors or keeping all guard ring pads floating, 3) removing all high voltage wires. There was no significant effect on the performance by action 1) or 2), while the noise was reduced by 60 % by action 3). Based on these observations, the 40MHz noise is related to the 40MHz clock activity in the circular buffer memory.

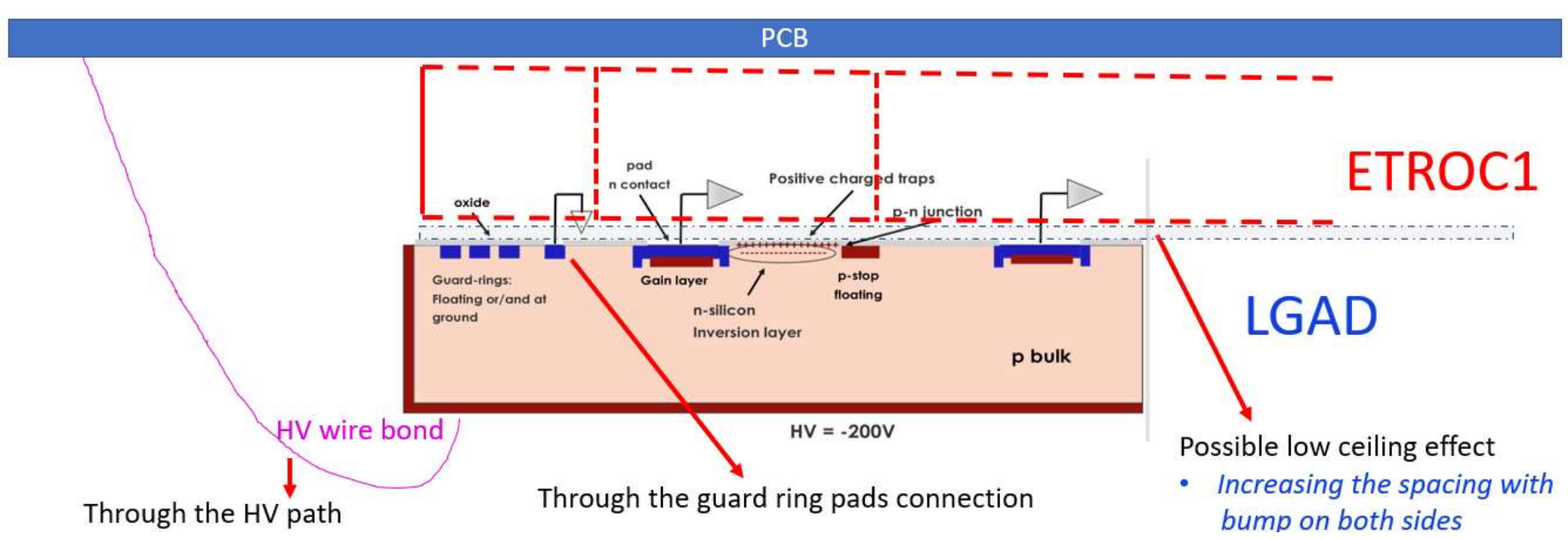

**Figure 15**. Diagram of bump-bonded chipset for 40 MHz noise investigation.

To address the noise issue in ETROC2, the following strategies have been considered and implemented, e.g., gating the memory clock based on hits by activating the memory clock only on valid TDC hits, offsetting some clocks deliberately to stagger clock transitions, putting a shielding layer on the top of ETROC2, separating 40 MHz clocks for readout and for TDC, and having the flexibility to disable readout clock for specified pixels. Additionally, system-level simulations and analysis have been performed to help verify the effectiveness of these measures in reducing noise and improving the overall performance of the system.

## 4. Conclusion

ETROC1 is the first full chain precision timing prototype ASIC for the CMS MTD endcap timing layer, aiming to study and demonstrate the performance of the full signal processing chain, with the goal to achieve 40 – 50 ps time resolution per hit. ETROC1 has a 5 × 5-pixel array with a 4 × 4 H-tree shaped clock distribution network that is scalable to the final full size of 16 × 16. The periphery circuits provide clock signals, readout functionalities, as well as standalone pixel readout and TDC blocks for testing purposes.

ETROC1 has been tested extensively under various stimuli with charge injection, laser, and beam. The time resolution of bare ETROC1 chips has achieved around 10 ps from on-pixel charge injection feature. Bump-bonded ETROC1 + LGAD chipsets have achieved around 20 ps resolution in laser test with large signal amplitudes. Their resolution in beam test has reached 41 – 43 ps per hit.

Noise related to 40 MHz clock was observed in bump-bonded chipsets. Investigations have shown that the noise is related to 40MHz clock activity in the circular buffer memory that is coupled through LGAD sensor. Strategies have been proposed and implemented in the design of ETROC2 and test boards to minimize the noise.

## Acknowledgments

This work has been authored by Fermi Research Alliance, LLC under Contract No. DE-AC02-07CH11359 with the US Department of Energy, Office of Science, Office of High Energy Physics.